# Topological Diagnosis of Optical Composites via Inversion of Nonlinear Dielectric Mixing Rules

*Proity Nayeeb Akbar*

*Department of Physics, Wesleyan University, 265 Church Street, Middletown, CT 06459-0155, USA*

*Email: pakbar@wesleyan.edu*

## Abstract

Accurate determination of the complex effective permittivity is fundamental to optical material engineering, but it remains a critical metrology challenge for heterogeneous systems. In polymer blends and optical composites, scattering and nonlinear dielectric effects severely distort spectral signatures, causing conventional linear unmixing and data-driven approaches to fail. Here, we present an inverse reconstruction framework that retrieves the broadband complex permittivity and constituent composition of strongly scattering mixtures directly from a single infrared (IR) extinction spectrum. The method integrates scattering theory, Lorentz oscillator modeling, and a generalized set of nonlinear effective medium approximations (EMAs) to simultaneously identify component spectra, estimate volume fractions, and, crucially, diagnose the underlying microstructure. The reconstruction algorithm demonstrates robust performance across synthetic two- and multi-component polymer blends, rigorously testing the efficacy of inverted, logarithmic, and cubic mixing regimes. By comparing the statistical causality and fitting quality of these competing EMAs, the framework uniquely provides a non-destructive, optical diagnosis of the blend's dominant interaction topology (e.g. co-continuous vs. stratified/series). The reconstructed permittivity spectra are dispersion-consistent and reveal physically interpretable optical properties across the full IR range. This framework establishes a new paradigm for inverse metrology in photonics, providing a necessary, physics-grounded foundation for the quantitative characterization and rational design of nonlinear optical composites. Specifically, by providing scattering-immune effective permittivity for forward modeling and delivering a physics-based diagnosis of the underlying microstructure, the framework enables engineers to reliably link fabrication parameters to the intended optical function.

## Introduction

Complex permittivity (or complex refractive index) is the central quantity in photonics and optical material engineering, governing emergent electromagnetic behavior in heterogeneous materials – from biological systems to advanced optical composites [1, 2, 3, 4]. While critical for the forward design and modeling of materials, like metamaterials and novel sensors, retrieving this intrinsic, broadband function directly from experimental data is notoriously difficult [5, 6]. Specifically, in the infrared (IR) regime, the mixing of components leads to strong scattering and nonlinear dielectric interactions [7, 8, 9], corrupting the measured absorption signature and rendering traditional analytical methods – such as the Beer-Lambert law [10] or Kramers-Kronig transformation [11] – inaccurate or inapplicable [12, 13, 14].

Recent advances in hyperspectral and spectroscopic imaging have created a demand for quantitative, spatially resolved chemical analysis [15, 16, 17, 18, 19, 20]. However, the data interpretation is critically bottlenecked by the spectral unmixing (SU) problem [21]. Most established SU algorithms rely on the Linear Mixing Model (LMM), which fundamentally assumes measured spectra are simple volume-fraction-weighted sums of pure components [22, 23, 24]. This assumption fails in scattering media where near-field coupling and intimate dielectric interactions cause spectral features to shift, broaden, and interact nonlinearly, resulting in data that lie on complex, nonlinear manifolds [25, 26, 9, 27]. In this context, "nonlinear" refers to the non-additive, geometry-dependent nature of the effective medium approximations (EMAs) used to describe the composite's permittivity. For these realistic systems, the inverse problem remains ill-posed: one must infer not only the component spectra and abundances but also the unknown physical interaction mechanism from minimal, noisy data [28, 29, 30, 31, 21].

Here, we present a fundamental advancement of the physics-informed inverse reconstruction framework [32, 33, 34], elevating its function from basic spectral deconvolution to broadband dielectric metrology and microstructure diagnosis. Our method rigorously integrates the physical process of light scattering with the constitutive physics of mixing (EMAs). The two-stage framework simultaneously: **1)** reconstructs the complex effective permittivity of the medium, a geometry-independent material constant required for forward modeling and design applications; **2)** recovers the pure component permittivity spectra (deconvolved from scattering); **3)** quantifies the component volume fractions; and **4)** diagnoses the internal microstructure, which is the primary determinant of the macroscopic optical performance of the composite. Crucially, we move beyond our prior work [32] restricted to simple or linear EMAs by explicitly incorporating a suite of nonlinear dielectric mixing rules, including the "inverted", "logarithmic", and "cubic" regimes [35, 36, 37].

The key to this diagnostic power lies in the distinct microscopic justifications of these EMAs. The theoretical foundation for relating component permittivity to effective permittivity lies in effective medium theory (EMT), defined by fundamental models, such as Maxwell-Garnett (matrix-inclusion topology) and Bruggeman (symmetric co-continuous topology) approximations [38, 37]. These are derived from long-wavelength field-averaging theories [38, 37].

To rigorously model the highly nonlinear dielectric effects typical of polymer blends, we utilize generalized EMT models that extend these foundations by spanning three distinct physical mixing scenarios. The "cubic" (Looyenga) model serves as a robust generalization of the Bruggeman approach, diagnosing a co-continuous network topology across wide volume fractions [39, 37]. The "inverted" (lower Wiener bound) model rigorously tests the stratified topology by representing the theoretical maximum electrical resistance (series mixing) [40, 41, 37]. Finally, the "logarithmic" (Lichtenecker) model is included to diagnose a statistical mixture of random grains, representing the case where the mixture's properties are determined by a volume average of the logarithms of the component properties [42, 37].

This systematic comparison of physically justified topologies is the core of our microstructure diagnosis, fundamentally refuting the notion that these models are heuristic. By testing which model yields the most statistically stable and physically causal complex permittivity spectra, the inversion framework provides a quantitative, non-destructive optical diagnosis of the material's underlying dielectric microstructure.

We demonstrate the framework's robustness across synthetic two- and multi-component polymer blends (representative of microstructured optical composites). These systems were specifically designed to yield micron-scale, free-standing particles — a geometry where Mie scattering theory [43] is the rigorous, first-principles model for describing the measured extinction cross-section [8, 44]. We intentionally utilize the same benchmark polymer systems established in the foundational work [32] to allow for a direct, quantitative comparison that verifies the metrology consistency of the effective permittivity recovery, while isolating the genuinely new physical contribution – the rigorous diagnosis of microstructure via nonlinear EMA comparison.

This work establishes a model-consistent, single-spectrum paradigm that automatically solves the nonlinear SU problem by grounding the unmixing in first-principles physics. It provides a computational foundation for quantitative hyperspectral imaging, enabling the design, characterization, and verification of nonlinear composites and functional photonic materials [45, 46, 47, 48, 49].

## Methods

This study presents a physics-based inverse reconstruction framework for recovering the complex effective permittivity and constituent composition of nonlinear dielectric mixtures from a single IR extinction spectrum. This method establishes a model-consistent inverse metrology standard that resolves component-specific contributions in strongly scattering and compositionally complex media. **Figure 1** provides a schematic overview of the complete reconstruction framework.

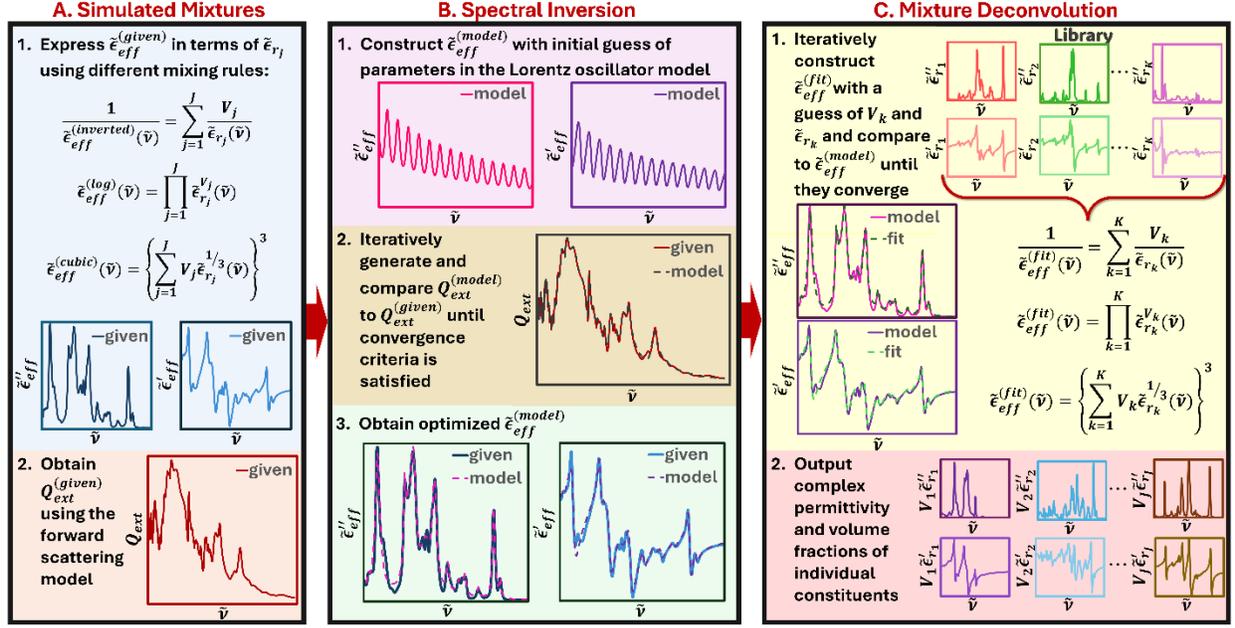

*Figure 1: Schematic of the Two-Stage Inverse Metrology Framework for Microstructure Diagnosis.* The framework rigorously integrates the physical process of light scattering with the constitutive physics of mixing to solve the inverse problem. **A. Simulated Mixtures (Stage A: Forward Model Simulation): 1)** Complex effective permittivity, $\tilde{\epsilon}_{eff}^{(given)}$, is constructed by applying three physically distinct mixing topologies – inverted ($\tilde{\epsilon}_{eff}^{(inverted)}$, representing the stratified/series topology), logarithmic ($\tilde{\epsilon}_{eff}^{(\log)}$, representing a statistical/random grain mixture), and cubic ($\tilde{\epsilon}_{eff}^{(cubic)}$, representing a co-continuous mixture) – to the constituent complex permittivity, $\tilde{\epsilon}_{r_j}$ and volume fraction, $V_j$. **2)** The 'ground truth' experimental extinction efficiency, $Q_{ext}^{(given)}$, is synthetically generated by solving Maxwell's equations (Mie theory) for $\tilde{\epsilon}_{eff}^{(given)}$. This synthetic generation is necessary to precisely validate the recovery of volume fractions and the diagnosis of the underlying mixing topology (inverted, logarithmic, or cubic) against an absolute known standard. **B. Spectral Inversion (Stage B: Effective Permittivity Recovery): 1)** The inverse metrology begins with spectral inversion, fitting the experimental extinction efficiency, $Q_{ext}^{(given)}$, using a Lorentz oscillator model (LOM) whose parameters describe the reconstructed permittivity, $\tilde{\epsilon}_{eff}^{(model)}$. **2)** The LOM-based extinction efficiency, $Q_{ext}^{(model)}$, is generated via the Mie model and the residual error is minimized yielding the scattering-immune, broadband material constant, $\tilde{\epsilon}_{eff}^{(model)}$, which is required for forward optical design. **C. Mixture Deconvolution (Stage C: Microstructure Diagnosis): 1)** The recovered $\tilde{\epsilon}_{eff}^{(model)}$ is simultaneously deconvolved using all three competing mixing rules to obtain the predicted spectrum, $\tilde{\epsilon}_{eff}^{(fit)}$. **2)** The optimization minimizes the residual error between $\tilde{\epsilon}_{eff}^{(model)}$ and $\tilde{\epsilon}_{eff}^{(fit)}$, isolating the component properties ($\tilde{\epsilon}_{r_k}$, $V_k$) and, most importantly, identifying the mixing topology (the EMA model) that yields the lowest error, providing the final microstructure diagnosis of the blend.

## Materials and Data

A collection of six organic polymers was used, such as polymethylmethacrylate (PMMA), polycarbonate (PC), polydimethylsiloxane (PDMS), polyetherimide (PEI), polyethylene terephthalate (PET), and polystyrene (PS), with distinct IR signatures derived from FTIR and ellipsometry measurements [50, 51, 52, 53, 54, 55]. These served as the ground truth for simulating mixtures with known composition and nonlinear interactions.

## Physical Justification of the Scattering Regime

The light-matter interaction in this framework is modeled using Mie scattering theory, which provides a rigorous solution to Maxwell's equations for spherical particles. The choice of this regime is dictated by the dimensionless size parameter, $x$, defined as:

$$x = 2\pi r \tilde{n}/\lambda, \tag{1}$$

where $r$ is the particle radius, $\tilde{n}$ is the refractive index of the surrounding medium, and $\lambda$ is the wavelength of the incident radiation. In the mid-infrared (mid-IR) range, $\lambda$ is approximately $2.5 - 20\ \mu m$ ($500 - 4000\ cm^{-1}$ in wavenumbers) and given the inclusion radius of $r = 5\ \mu m$ used in this study, $x$ ranges approximately from 1.5 to 12.5. This range corresponds squarely to the Mie scattering regime ($x \approx 1$), where the wavelength is comparable to the inclusion size. In this regime, the Rayleigh approximation is physically insufficient as it neglects the internal field distributions and phase shifts that occur within the microstructured inclusions.

While the current validation utilizes spherical geometries to isolate the performance of the Microstructure Diagnosis algorithm (Stage C of **Figure 1**), the modular architecture of the Stage B inversion pipeline is designed for flexibility. Future applications involving non-spherical or highly anisotropic morphologies can be accommodated by substituting the Mie forward model with alternative solvers, such as the T-matrix method for non-spherical particles or Fresnel-based models for stratified media.

## Simulated Mixtures (Stage A: Defining Scattering Model)

Synthetic mixtures were generated (stage A of **Figure 1**) by combining polymers in controlled volume fractions under three nonlinear mixing models: inverted, logarithmic, and cubic [35, 36, 37]. The deliberate comparison of these three models, which represent distinct physical topologies (series/stratified, statistical/random grain, and co-continuous/symmetric, respectively), is the foundation of our microstructure diagnosis. As outlined below, equations (2) to (4) represent the complex effective permittivity, $\tilde{\epsilon}_{eff}^{(inverted)}$, $\tilde{\epsilon}_{eff}^{(log)}$, and $\tilde{\epsilon}_{eff}^{(cubic)}$, obtained using the inverted, logarithmic, and cubic mixing rule, respectively.

$$\frac{1}{\tilde{\epsilon}_{eff}^{(inverted)}(\tilde{v})} = \frac{1}{\tilde{\epsilon}_{eff}^{\prime(inverted)}(\tilde{v}) + i\tilde{\epsilon}_{eff}^{\prime\prime(inverted)}(\tilde{v})} = \sum_{j=1}^{J} \frac{V_j}{\tilde{\epsilon}_{r_j}(\tilde{v})}, \tag{2}$$

$$\tilde{\epsilon}_{eff}^{(log)}(\tilde{v}) = \tilde{\epsilon}'^{(log)}_{eff}(\tilde{v}) + i\tilde{\epsilon}''^{(log)}_{eff}(\tilde{v}) = \prod_{j=1}^{J} \tilde{\epsilon}_{r_j}^{V_j}(\tilde{v}), \quad (3)$$

$$\tilde{\epsilon}_{eff}^{(cubic)}(\tilde{v}) = \tilde{\epsilon}'^{(cubic)}_{eff}(\tilde{v}) + i\tilde{\epsilon}''^{(cubic)}_{eff}(\tilde{v}) = \left\{ \sum_{j=1}^{J} V_j \tilde{\epsilon}_{r_j}^{\frac{1}{3}}(\tilde{v}) \right\}^{3}. \quad (4)$$

The quantities $\tilde{\epsilon}'^{(inverted)}_{eff}$, $\tilde{\epsilon}''^{(inverted)}_{eff}$, $\tilde{\epsilon}'^{(log)}_{eff}$, $\tilde{\epsilon}''^{(log)}_{eff}$, $\tilde{\epsilon}'^{(cubic)}_{eff}$, and $\tilde{\epsilon}''^{(cubic)}_{eff}$ express the real and imaginary components of $\tilde{\epsilon}_{eff}^{(inverted)}$, $\tilde{\epsilon}_{eff}^{(log)}$, and $\tilde{\epsilon}_{eff}^{(cubic)}$, respectively. In this expression, $\tilde{\epsilon}_{r_j}$ and $V_j$ denote the complex permittivity and volume fraction of the $j$-th polymer component in the composite system. The index $j$ runs over the individual components of the mixture, with the total number of components given by $J$.

Each mixture was then modeled as a homogeneous sphere of radius $5\ \mu m$ and its extinction efficiency, $Q_{ext}$, computed using the PyMieScatt package [56], which solves Maxwell's equations for a given complex effective permittivity, $\tilde{\epsilon}_{eff}^{(given)}$. Since the extinction efficiency conveys the same information about the missing light due to the shape and chemical composition of the sample as the apparent IR absorbance, $A$ [8, 44], this data is utilized for reconstruction and referred to as the experimental extinction spectrum, $Q_{ext}^{(given)}$. The associated complex effective permittivity, $\tilde{\epsilon}_{eff}^{(given)}$, corresponding to $Q_{ext}^{(given)}$ can be expressed by the complex effective permittivity obtained from any of the three mixing models: $\tilde{\epsilon}_{eff}^{(inverted)}$, $\tilde{\epsilon}_{eff}^{(log)}$, or $\tilde{\epsilon}_{eff}^{(cubic)}$.

Synthetically generating 'ground truth' extinction spectra facilitates a rigorous benchmark of the inverse algorithm by decoupling its performance from experimental uncertainties such as sample preparation variance and instrument noise. By using the experimental values of the constituent refractive indices of the six organic polymers (PMMA, PC, PDMS, PEI, PET, and PS) as the basis for Mie scattering calculations (Stage A of **Figure 1**), we ensure that the simulated 'experimental' spectra maintain high physical fidelity because they incorporate the full experimental dispersion profiles of the pure constituents, providing a known target for the precise evaluation of volume fraction recovery and topological diagnosis.

## Inverse Metrology Pipeline

The inverse reconstruction framework involves two stages:

### Spectral Inversion (Stage B: Effective Permittivity Recovery)

Spectral inversion (stage B of **Figure 1**) accounts for scattering and corrects distortions to isolate the underlying dielectric response. Using a Lorentz oscillator model [43, 34, 33, 32, 57, 58], we fit the experimental extinction spectrum, $Q_{ext}^{(given)}$, by adjusting a minimal set of parameters, $\epsilon_\infty$, $\tilde{v}_{0_{eff}}^{(n)}$, $\tilde{v}_{p_{eff}}^{(n)}$, and $\gamma_{eff}^{(n)}$, to recover the real, $\tilde{\epsilon}'^{(model)}_{eff}$, and imaginary, $\tilde{\epsilon}''^{(model)}_{eff}$, components of the mixture's complex effective permittivity, $\tilde{\epsilon}_{eff}^{(model)}$. The index $n = 1, \dots, N$ represents the different vibrational resonances within the mixture. These parameters describe the shape and internal structure of the modeled homogeneous mixed-composition sphere, which exhibits nonlinear interactions between its components. For further details on their physical significance, see the Supplementary Materials and our previous work [33, 32, 34].

$$\tilde{\epsilon}_{eff}^{(model)}(\tilde{v}) = \tilde{\epsilon}'^{(model)}_{eff}(\tilde{v}) + i\tilde{\epsilon}''^{(model)}_{eff}(\tilde{v}), \tag{5}$$

$$\tilde{\epsilon}'^{(model)}_{eff}(\tilde{v}) = \sum_{n=1}^{N} \left\{ \epsilon_{\infty} + \frac{\tilde{v}_{p_{eff}}^{(n)\,2} \left( \tilde{v}_{0_{eff}}^{(n)\,2} - \tilde{v}^2 \right)}{\left( \tilde{v}_{0_{eff}}^{(n)\,2} - \tilde{v}^2 \right)^2 + \left( \tilde{v}\gamma_{eff}^{(n)} \right)^2} \right\}, \tag{6}$$

$$\tilde{\epsilon}''^{(model)}_{eff}(\tilde{v}) = \sum_{n=1}^{N} \left\{ \frac{\tilde{v}_{p_{eff}}^{(n)\,2} \tilde{v}\gamma_{eff}^{(n)}}{\left( \tilde{v}_{0_{eff}}^{(n)\,2} - \tilde{v}^2 \right)^2 + \left( \tilde{v}\gamma_{eff}^{(n)} \right)^2} \right\}. \tag{7}$$

The extinction efficiency corresponding to $\tilde{\epsilon}_{eff}^{(model)}$, denoted $Q_{ext}^{(model)}$, is calculated by solving Maxwell's equations using the PyMieScatt package [56]. The optimal match to $Q_{ext}^{(given)}$ is then obtained by minimizing the residual error, $S$, between the two functions, evaluated over a set of wavenumbers, $\tilde{v}_l$, spanning the mid-IR spectrum (8).

$$S\left\{\epsilon_{\infty}, \tilde{v}_{0_{eff}}^{(n)}, \tilde{v}_{p_{eff}}^{(n)}, \gamma_{eff}^{(n)}\right\} = \sum_{l} \left\{ Q_{ext}^{(model)}\left[\left\{\epsilon_{\infty}, \tilde{v}_{0_{eff}}^{(n)}, \tilde{v}_{p_{eff}}^{(n)}, \gamma_{eff}^{(n)}\right\}, \tilde{v}_l\right] - Q_{ext}^{(given)}(\tilde{v}_l) \right\}^2, \tag{8}$$

## Mixture Deconvolution (Stage C: Microstructure Diagnosis)

The reconstructed complex permittivity, $\tilde{\epsilon}_{eff}^{(model)}$, is then decomposed into contributions from individual components (stage C of **Figure 1**). The algorithm explores combinations of candidate materials' complex permittivities, $\tilde{\epsilon}_{r_k}$, and corresponding volume fractions, $V_k$, using either a grid search [59] for low-dimensional problems or gradient descent [60] for high-dimensional mixtures, where $k = 1, \ldots, K$ indexes materials in the spectral database. The optimization is performed for each of the three mixing models ($\tilde{\epsilon}_{eff}^{(inverted)}$, $\tilde{\epsilon}_{eff}^{(log)}$, and $\tilde{\epsilon}_{eff}^{(cubic)}$) independently.

Each combination produces a predicted spectrum for the complex effective permittivity, and the model that yields the lowest residual sum of squares (9) and the most physically causal complex permittivity spectrum is designated as the fitted spectrum, $\tilde{\epsilon}_{eff}^{(fit)}$, and provides the diagnostic conclusion on the blend's topology.

$$RSS[\tilde{\epsilon}_{r_k}, V_k] = \sum_{l} \left\{ \tilde{\epsilon}_{eff}^{(fit)}[\{\tilde{\epsilon}_{r_k}, V_k\}; \tilde{v}_l] - \tilde{\epsilon}_{eff}^{(model)}(\tilde{v}_l) \right\}^2. \tag{9}$$

Both stages are fully automated, require no prior assumptions about the number or identity of components in the mixture, and are robust to initial conditions (see Results and Supplementary Materials for details). Since each dielectric function is characterized by three parameters—$\tilde{v}_{0_{eff}}^{(n)}$, $\tilde{v}_{p_{eff}}^{(n)}$, and $\gamma_{eff}^{(n)}$—and the model includes a single DC offset, $\epsilon_{\infty}$, the optimization corresponds to solving an inverse scattering problem in a $91 - 103$ dimensional parameter space, depending on mixture complexity. For additional algorithmic and mathematical details, please see the Supplementary Material or refer to our previous work [33, 32, 34].

## Accuracy and Sensitivity Testing

The inverse reconstruction framework is validated across two- and multi-component mixtures using both known and randomized initial conditions. Performance was assessed based on reconstruction error in complex permittivity and predicted volume fractions.

## Results

The reconstruction framework was evaluated to establish its capability as a physics-informed inverse metrology tool by rigorously recovering the broadband complex permittivity, $\tilde{\epsilon}_{eff}^{(model)}$, and performing subsequent microstructure diagnosis based on the comparison of nonlinear mixing models. Our tests span two-component and multi-component mixed-composition spheres at different volume fractions governed by inverted, logarithmic, and cubic mixing rules.

## Spectral Inversion (Stage B: Effective Permittivity Recovery)

The spectral inversion (stage B of **Figure 1**) was first validated on simulated two- and multi-component polymer spheres by comparing reconstructed complex effective permittivity spectra against experimentally obtained ground truth values. For the two-component analysis, 15 mixture combinations were simulated using three nonlinear mixing models: inverted, logarithmic, and cubic. This extensive matrix of experiments was designed to verify the consistency and accuracy of the effective permittivity recovery across different chemical compositions, volume ratios, and known mixing topologies.

The mixtures were labeled as follows: PC-PDMS, PC-PEI, PC-PET, PEI-PDMS, PEI-PET, PET-PDMS, PMMA-PC, PMMA-PDMS, PMMA-PEI, PMMA-PET, PMMA-PS, PC-PS, PDMS-PS, PEI-PS, and PET-PS. Each mixture was tested at three distinct volume fractions: 10%: 90%, 30%: 70%, and 50%: 50%. The naming convention (e.g., PC-PDMS or PC-PEI) indicates the two polymers involved, with the first listed component appearing at the first volume fraction. For example, PC-PDMS at 10%: 90% refers to 10% polycarbonate (PC) and 90% polydimethylsiloxane (PDMS).

Multi-component systems were composed of PMMA, PC, PDMS, PET, PEI, and PS, and tested under six composition scenarios with varying dominant and minor components: **A)** 40%: 10%: 5%: 15%: 20%: 10%, **B)** 30%: 20%: 10%: 5%: 15%: 20%, **C)** 5%: 5%: 65%: 5%: 5%: 15%, **D)** 0%: 0%: 78%: 3%: 7%: 12%, **E)** 19%: 13%: 41%: 8%: 18%: 1%, and **F)** 16.5%: 16.5%: 16.5%: 16.5%: 16.5%: 17.5%. Here, a mixture labeled 19%: 13%: 41%: 8%: 18%: 1% PMMA-PC-PDMS-PET-PEI-PS represents a composite made up of 19% polymethylmethacrylate (PMMA), 13% polycarbonate (PC), 41% polydimethylsiloxane (PDMS), 8% polyethylene terephthalate (PET), 18% polyetherimide (PEI), and 1% polystyrene (PS).

The results of spectral inversion for two-component mixtures modeled using the inverted, cubic, and logarithmic mixing rules across the three volume fractions are shown in **Figures S1** through **S3**, respectively, in the Supplemental Materials section. **Figures 2** through **4** display the results of stage B for multi-component mixtures, providing the complex effective permittivity input for the diagnostic phase. We compare the reconstructed effective permittivity, $\tilde{\epsilon}_{eff}^{(model)}$, obtained when assuming each of the three mixing topologies: inverted (**Figure 2**), logarithmic (**Figure 3**), and cubic (**Figure 4**). Each figure contains three plots: panel **(a)** compares the imaginary part of the experimental complex effective permittivity, $\tilde{\epsilon}_{eff}''$ (solid yellow), with the reconstructed imaginary part, $\tilde{\epsilon}_{eff}''^{(model)}$ (dashed purple), at the specified volume fractions; panel **(b)** shows the real part of the experimental complex effective permittivity, $\tilde{\epsilon}_{eff}'$ (solid green), against the reconstructed real part, $\tilde{\epsilon}_{eff}'^{(model)}$ (dashed blue); and panel **(c)** compares the experimental extinction efficiency, $Q_{ext}^{(given)}$ (solid light blue), with the numerical fit, $Q_{ext}^{(model)}$ (dashed dark red), across the same volume fractions.

As shown in **Figures 2–4(c)**, across every tested scenario, the reconstructed extinction efficiency spectra (dashed dark red) demonstrate exceptional agreement with the synthetically generated ground-truth data (solid light blue), confirming the robust, scattering-immune capability of the inversion framework. Crucially, the reconstructed complex effective permittivity, $\tilde{\epsilon}_{eff}^{(model)}$ (both real and imaginary components), is physically consistent, causal, and free from geometric artifacts typically present in the raw extinction data. The utilization of synthetic data derived from experimental constituent indices ensures that any residual deviation in the fit is a result of the algorithm's performance rather than physical sample variance. This success in stage B confirms that the framework reliably extracts the intrinsic, broadband material constant required for optical modeling and design. This precise recovery of the complex effective permittivity across a broad spectrum of compositions, volume fractions, and known mixing topologies provides the necessary, accurate input for the subsequent stage C: microstructure diagnosis.

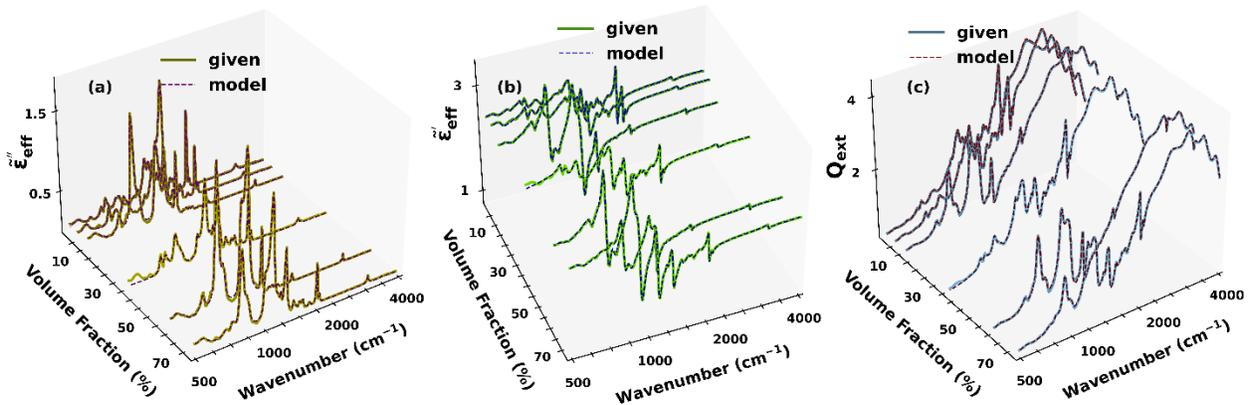

*Figure 2: Inverse Metrology of Multi-Component Blends: Broadband Effective Permittivity Recovery under the Inverted Mixing Topology.* Results from spectral inversion

(Stage B) for six multi-component mixtures (PMMA-PC-PDMS-PET-PEI-PS), at volume fractions: 40%: 10%: 5%: 15%: 20%: 10%, 30%: 20%: 10%: 5%: 15%: 20%, 5%: 5%: 65%: 5%: 5%: 15%, 0%: 0%: 78%: 3%: 7%: 12%, 19%: 13%: 41%: 8%: 18%: 1%, and 16.5%: 16.5%: 16.5%: 16.5%: 16.5%: 17.5%, assuming an inverted mixing topology (stratified/series). This figure validates the fidelity of the inverse reconstruction framework by demonstrating the accurate recovery of the complex effective permittivity ($\tilde{\epsilon}_{eff}^{(model)}$), the fundamental material constant. Panel **(a)** and **(b)** show the exceptional agreement between the experimental (ground truth) and reconstructed real ($\tilde{\epsilon}_{eff}^{\prime(model)}$) and imaginary ($\tilde{\epsilon}_{eff}^{\prime\prime(model)}$) components, confirming the Kramers-Kronig consistency of the recovered dielectric function. Panel **(c)** shows the tight fit of the extinction efficiency ($Q_{ext}^{(model)}$) confirming the full inversion of the scattering problem from a single spectrum. (The figures are plotted with respect to the volume fraction of PDMS for visualization.)

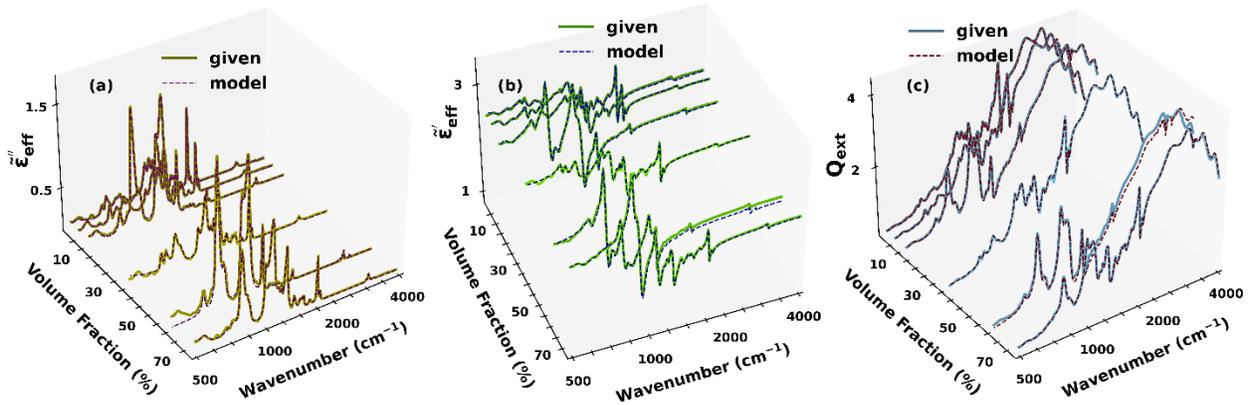

*Figure 3: Inverse Metrology of Multi-Component Blends: Broadband Effective Permittivity Recovery under the Logarithmic Mixing Topology.* Results from spectral inversion (Stage B) for six multi-component mixtures (PMMA-PC-PDMS-PET-PEI-PS), at volume fractions: 40%: 10%: 5%: 15%: 20%: 10%, 30%: 20%: 10%: 5%: 15%: 20%, 5%: 5%: 65%: 5%: 5%: 15%, 0%: 0%: 78%: 3%: 7%: 12%, 19%: 13%: 41%: 8%: 18%: 1%, and 16.5%: 16.5%: 16.5%: 16.5%: 16.5%: 17.5%, assuming a logarithmic mixing topology (statistical/random grain). This figure validates the fidelity of the inverse reconstruction framework by demonstrating the accurate recovery of the complex effective permittivity ($\tilde{\epsilon}_{eff}^{(model)}$), the fundamental material constant. Panel **(a)** and **(b)** show the exceptional agreement between the experimental (ground truth) and reconstructed real ($\tilde{\epsilon}_{eff}^{\prime(model)}$) and imaginary ($\tilde{\epsilon}_{eff}^{\prime\prime(model)}$) components, confirming the Kramers-Kronig consistency of the recovered dielectric function. Panel **(c)** shows the tight fit of the extinction efficiency ($Q_{ext}^{(model)}$) confirming the full inversion of the scattering problem from a single spectrum. (The figures are plotted with respect to the volume fraction of PDMS for visualization.)

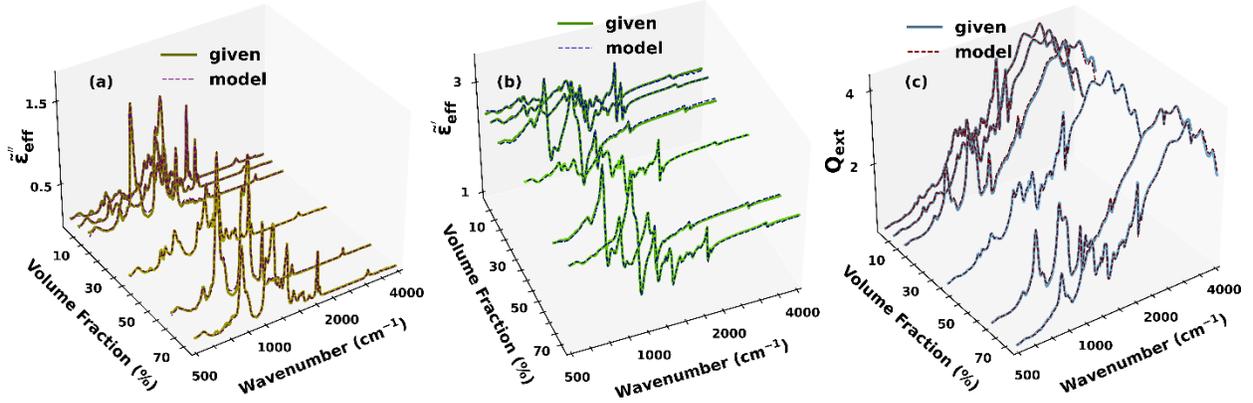

*Figure 4: Inverse Metrology of Multi-Component Blends: Broadband Effective Permittivity Recovery under the Cubic Mixing Topology. Results from spectral inversion (Stage B) for six multi-component mixtures (PMMA-PC-PDMS-PET-PEI-PS), at volume fractions: 40%: 10%: 5%: 15%: 20%: 10%, 30%: 20%: 10%: 5%: 15%: 20%, 5%: 5%: 65%: 5%: 5%: 15%, 0%: 0%: 78%: 3%: 7%: 12%, 19%: 13%: 41%: 8%: 18%: 1%, and 16.5%: 16.5%: 16.5%: 16.5%: 16.5%: 17.5%, assuming a cubic mixing topology (co-continuous/symmetric). This figure validates the fidelity of the inverse reconstruction framework by demonstrating the accurate recovery of the complex effective permittivity ($\tilde{\epsilon}_{eff}^{(model)}$), the fundamental material constant. Panel **(a)** and **(b)** show the exceptional agreement between the experimental (ground truth) and reconstructed real ($\tilde{\epsilon}_{eff}^{\prime(model)}$) and imaginary ($\tilde{\epsilon}_{eff}^{\prime\prime(model)}$) components, confirming the Kramers-Kronig consistency of the recovered dielectric function. Panel **(c)** shows the tight fit of the extinction efficiency ($Q_{ext}^{(model)}$) confirming the full inversion of the scattering problem from a single spectrum. (The figures are plotted with respect to the volume fraction of PDMS for visualization.)*

# Mixture Deconvolution (Stage C: Microstructure Diagnosis via Topological Comparison)

Following the successful recovery of the complex effective permittivity ($\tilde{\epsilon}_{eff}^{(model)}$) in stage B, the microstructure diagnosis (Stage C of **Figure 1**) framework was applied. The stage compares the physical fidelity of three competing mixing topologies (inverted, logarithmic, and cubic) to rigorously determine the dominant interaction mechanism governing the blend's optical response. The chosen model must yield the most statistically stable and physically causal permittivity spectra for each component in the mixture, simultaneously providing accurate volume fractions for each.

To rigorously test the diagnostic power of the framework, the $\tilde{\epsilon}_{eff}^{(model)}$ spectrum recovered in stage B was deconvolved independently using the inverted, logarithmic, and cubic mixing models. As demonstrated by the topological validation matrix (**Supplementary Tables S1** and **S2**), across all tested two- and multi-component synthetic systems, the model that minimized the residual sum of squares (RSS) and yielded the most physically causal permittivity spectra always coincided precisely with the mixing model used to generate the forward simulation data. This outcome is not a tautology, but a definitive validation of the framework's diagnostic fidelity. It proves that the inverse process can rigorously distinguish between the distinct physical assumptions (topologies) embedded in each EMA model. For instance, when the input effective permittivity, $\tilde{\epsilon}_{eff}^{(given)}$, was generated assuming a cubic (Looyenga) topology, the inverse framework's minimization process strongly selected the cubic model, thereby, correctly diagnosing a co-continuous network microstructure for the mixed-composition polymeric spheres. Conversely, the systematic rejection of the inverted model (series/stratified topology) and the logarithmic model (statistical/random grain topology) confirms that the blend's effective permittivity is governed by uniform interpenetration, rather than barrier resistance or pure statistical averaging of non-interacting particles. This topological diagnosis is the primary diagnostic output of the reconstruction framework.

To contextualize the necessity of the nonlinear framework, a comparative analysis with the Linear Mixing Model (LMM) utilized in previous work [32] was performed. As detailed in **Supplementary Tables S1** and **S2**, the LMM consistently yields significantly higher residuals when applied to microstructured composites in the Mie scattering regime. Physically, the LMM failure stems from its inability to account for the phase-shifted extinction and localized resonance shifts inherent in scattering media. Unlike the nonlinear topologies investigated here, the LMM assumes non-interacting components, which fails to satisfy Kramers-Kronig consistency in the presence of strong scattering. This leads to non-causal reconstructions of effective permittivity, whereas the nonlinear manifold successfully unmixes these dielectric interactions to recover the intrinsic material properties.

With the correct mixing topology successfully diagnosed, the framework's analytical performance was assessed. The accuracy of component quantification serves as secondary validation, demonstrating that the chosen EMA model not only describes the underlying physics but also provides reliable numerical results.

In two-component systems, both grid-search and gradient descent yielded accurate and consistent results, with volume fraction predictions deviating less than ~2% from ground truth on average. Results for PC-PDMS, PC-PET, and PEI-PET are presented in **Figure 5**; reconstructions for the remaining mixtures are summarized in **Supplementary Table S3**. The near-identical output from both algorithms highlights the reliability of the method in identifying components and quantifying their concentrations — without requiring prior knowledge of the materials, their proportions, or the mixing rule. Furthermore, for multi-component systems (up to six components), gradient descent successfully identified all component identities and recovered their volume fractions across all tested mixing models (**Figure 6, Supplementary Table S4**). Dominant species, such as PDMS, were reconstructed with standard deviations under 0.5%, demonstrating that the two-stage reconstruction framework, once the topology is known, remains robust even for low abundance components like PS, where variance was only slightly higher.

Although the framework is designed to operate without prior assumptions, it can readily incorporate partial knowledge of component identities, concentrations, or mixing behavior. Supplying such information can further improve reconstruction accuracy and reduce computational cost — especially in targeted diagnostic or materials characterization workflows.

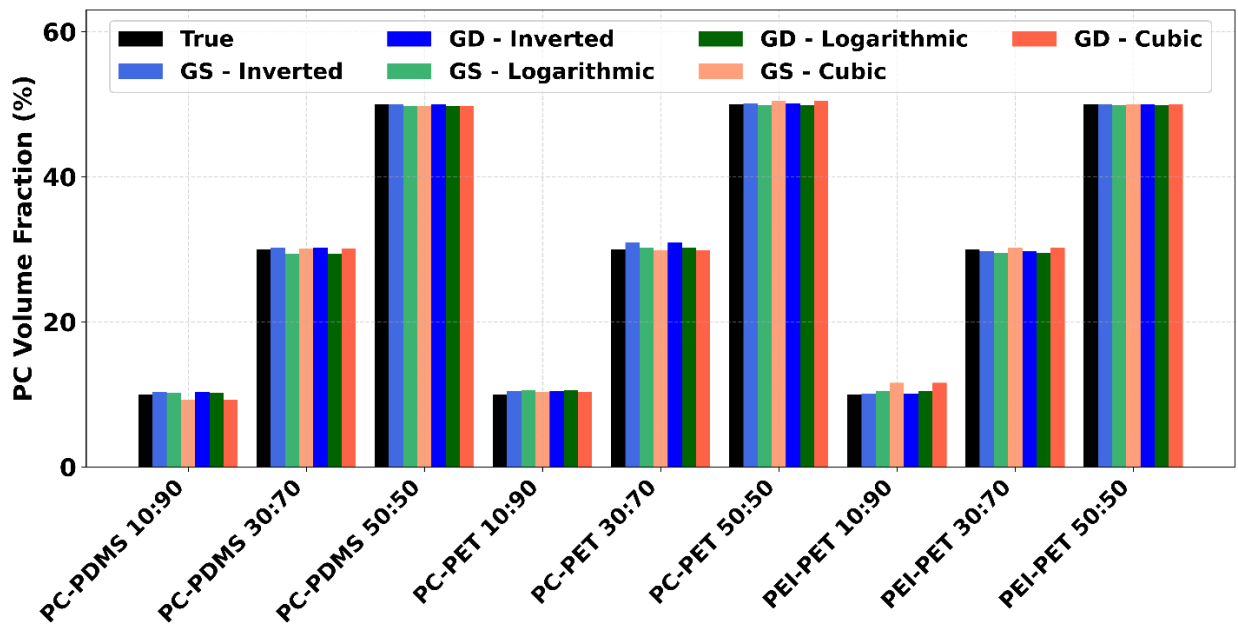

*Figure 5: Validation of Analytical Accuracy for Two-Component Systems. Mixture deconvolution results using grid search (GS) and gradient descent (GD), generated by the three competing topological models (inverted, logarithmic, and cubic), showing the accuracy of component identification and volume fraction recovery for two-component polymer blends (PC-PDMS, PC-PET, and PEI-PET) at three volume fractions (10%: 90%, 30%: 70%, and 50%: 50%). The consistent, high accuracy of the recovered volume fractions validates the fidelity of the selected, physically diagnosed mixing topology. While volume fraction recovery is stable across all models, microstructure diagnosis is determined by the RSS values in **Table S1**.*

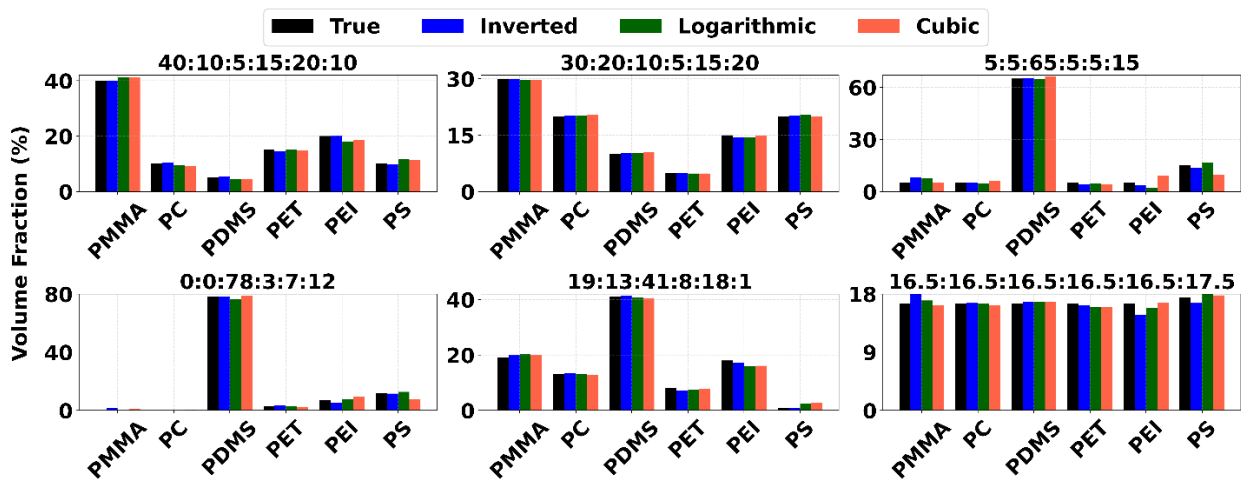

*Figure 6: Validation of Analytical Accuracy for Multi-Component Systems. Mixture deconvolution results using gradient descent, generated by the three competing topological models (inverted, logarithmic, and cubic), showing the accuracy of component identification and volume fraction recovery for multi-component polymer blends (PMMA-PC-PDMS-PET-PEI-PS) at six volume fractions (40%: 10%: 5%: 15%: 20%: 10%, 30%: 20%: 10%: 5%: 15%: 20%, 5%: 5%: 65%: 5%: 5%: 15%, 0%: 0%: 78%: 3%: 7%: 12%, 19%: 13%: 41%: 8%: 18%: 1%, and 16.5%: 16.5%: 16.5%: 16.5%: 16.5%: 17.5%). The consistently low error in volume fraction recovery confirms the predictive accuracy of the reconstruction framework across different composition and topological assumptions. While volume fraction recovery is stable across all models, microstructure diagnosis is determined by the RSS values in **Table S2**.*

## Sensitivity to Initial Conditions

Nonlinear inversion algorithms can be highly sensitive to initial conditions, posing a significant challenge for automated characterization. To evaluate the stability and reliability of our inverse metrology framework, we performed 2000 independent reconstructions for selected two-component (30%: 70% PC-PDMS) and multi-component (19%: 13%: 41%: 8%: 18%: 1% PMMA-PC-PDMS-PET-PEI-PS) mixtures under randomized initial conditions. This test verifies that the optimization landscape is well-behaved and that the framework yields consistent results regardless of the starting point.

The sensitivity analysis was conducted across all three competing mixing topologies (inverted, logarithmic, and cubic) using both grid search and gradient descent algorithms (**Figures S4-S6**). In each case, panel **(a)** presents the results from the grid search, while panel **(b)** shows the outcomes from gradient descent. The predicted volume fractions for the two-component 30%: 70% PC–PDMS mixtures (**Figures S4-S6**) consistently resulted in a narrow distribution clustered tightly around the ground truth value and a right-skewed histogram, indicating deviations from normality.

Under the inverted mixing rule (**Figure S4**), grid search predictions primarily fall between 29% and 33% with a mean of 30.7%, median of 30.5%, mode of 30.2% and a standard deviation of 0.85%. This reflects a reconstruction error of approximately 0.7%. Gradient descent results for the same case fall between 29.5% and 31.5%, with a mean of 30.1%, median of 30.1%, mode of 25.0%, and a slightly narrower standard deviation of 0.65%, indicating slightly improved precision. Under the logarithmic mixing rule (**Figure S5**), the grid search yields values between 28.5% and 31%, with a mean and median of 29.6%, mode of 29.6%, and standard deviation of 0.35%, leading to a small reconstruction error of 0.4%. The gradient descent algorithm produces similar results — a mean and median of 29.6%, mode of 30.0%, and standard deviation of 0.35% — also reflecting a 0.4% deviation from the true value. Under the cubic mixing rule (**Figure S6**), grid search predictions range from 28% to 31.2% with a mean of 29.4%, median and mode of 29.3%, and a standard deviation of 0.41%, corresponding to a 0.6% reconstruction error. Gradient descent estimates fall between 28.5% and 30.7%, with a mean and median of 29.3%, mode of 30.0%, and a slightly larger standard deviation of 0.70%, resulting in a 0.7% deviation.

Across all three mixing rules and both algorithms, the mean and median predicted volume fractions deviate by less than 1% from the true value. Specifically, the standard deviation remained low, indicating high precision and low variability in the optimization process regardless of the assumed EMA topology. This consistency across fundamentally distinct physical models confirms that the local minima found by the search algorithms are highly stable.

The predicted volume fractions for each component in the multi-component mixture, obtained through gradient descent, are shown in **Figures 7-9**, for the inverted, logarithmic, and cubic mixing rules, respectively. In each case, panel **(a)** displays an overview of the volume fraction histograms for all components in the multi-component mixture, plotted together in a single frame across the full spectrum. Panel **(b)** shows the volume fraction distributions of the individual components separately, with each component presented in its own frame.

For the inverted mixture interaction (**Figure 7**), the volume fractions are primarily distributed as follows: PMMA (16.5% − 21.5%, median 19.6%), PC (10% − 14%, median 12.3%), PDMS (39% − 41%, median 40.8%), PET (3% − 12%, median 7.9%), PEI (14% − 21%, median 18.3%), and PS (0% − 4.5%, median 1.1%). The predicted means are approximately 19.5 (PMMA), 12.1% (PC), 40.7% (PDMS), 8.0% (PET), 18.0% (PEI), and 1.4% (PS), with corresponding standard deviations of 1.0%, 0.8%, 0.4%, 1.5%, 1.2%, and 1.0%, respectively. Under the logarithmic interactions (**Figure 8**), the volume fraction ranges are similar: PMMA (16.5 − 24%, median 21.5%), PC (10% − 14%, median 12.0%), PDMS (39.5% − 42%, median 40.8%), PET (6% − 11%, median 8.1%), PEI (13% − 20%, median 16.6%), and PS (0% − 5%, median 1.7%). The corresponding means are 21.1% (PMMA), 11.9% (PC), 40.8% (PDMS), 8.3% (PET), 16.7% (PEI), and 1.8% (PS) with standard deviations of 1.7%, 0.8%, 0.4%, 1.0%, 1.3%, and 1.2%. Results for the cubic mixing rule (**Figure 9**) show nearly identical ranges: PMMA (16.5% − 24%, median 20.6%), PC (10% − 14%, median 12.1%), PDMS (39.5% − 42%, median 40.7%), PET (6% − 11%, median 8.0%), PEI (13% − 20%, median 16.8%), and PS (0% − 5%, median 1.9%). The predicted means are 21.0% (PMMA), 11.9% (PC), 40.8% (PDMS), 8.2% (PET), 16.8% (PEI), and 2% (PS), with standard deviations of 1.9%, 0.8%, 0.5%, 1.0%, 1.4%, and 1.2%, respectively.

We draw two key inferences from this multi-component stability analysis: **1)** topological insensitivity and **2)** high robustness/stability. Firstly, the range, mean, and median volume fractions predicted under the logarithmic and cubic models are very similar and differ only slightly from those obtained under the inverted rule. This indicates the robustness of the entire metrology approach, namely, that the gradient descent algorithm consistently delivers accurate predictions irrespective of the specific mixing topology assumed. This result strongly reinforces the reliability of the topological diagnosis performed in stage C. Secondly, the reconstruction of the most dominant component, PDMS, remains highly stable (reconstructed volume fraction ∼ 40.7% − 40.8%) across all mixing rules, exhibiting the lowest variability (standard deviation ∼ 0.4% − 0.5%). In contrast, the largest variability is observed for PS, the component with the smallest contribution to the mixture. The relatively large standard deviation for PS (compared to its mean) highlights the inherent physical limitation of resolving components present at very low concentrations, which is anticipated given the minimal spectral signature of minor constituents.

These results confirm that the optimization landscape is sufficiently smooth and well-behaved, reinforcing the suitability of the reconstruction framework for automated inverse modeling and microstructure diagnosis in practical optical materials characterization. However, for significantly more complex systems — such as mixtures with an even higher number of components, substantial spectral overlap, or low signal-to-noise ratios — it is acknowledged that sensitivity to initial conditions could become more pronounced. That said, the approach developed in this study is still broadly applicable to a wide range of dielectric materials. Most dielectric materials exhibit characteristic absorption features in the infrared region, and while peak positions and widths may vary, the general structure of their IR spectra tends to follow similar vibrational modes [28]. Therefore, the methodology can be robustly extended to a wide range of dielectric materials characterized by strong absorption features in the infrared region.

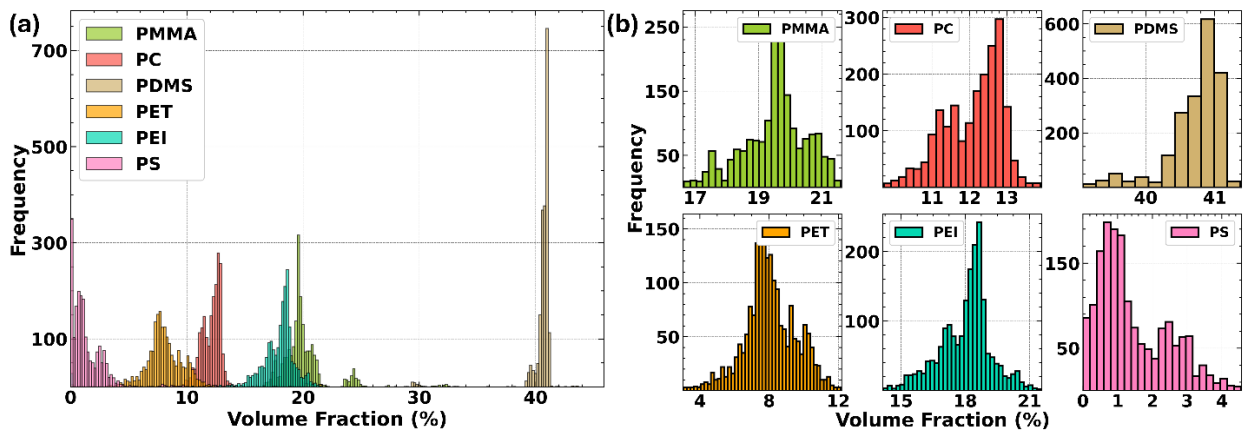

*Figure 7: Algorithm Stability and Precision under the Inverted Mixing Topology.* *Results from the sensitivity study for the multi-component mixture (19%: 13%: 41%: 8%: 18%: 1% PMMA-PC-PDMS-PET-PEI-PS) using gradient descent across 2000 randomized initial conditions. The inverted mixing rule (representing stratified/series topology) is applied for the volume fraction reconstruction. Panel* **(a)** *shows combined histograms of all components across the full volume fraction spectrum and panel* **(b)** *illustrates the individual component distributions shown in separate frames.*

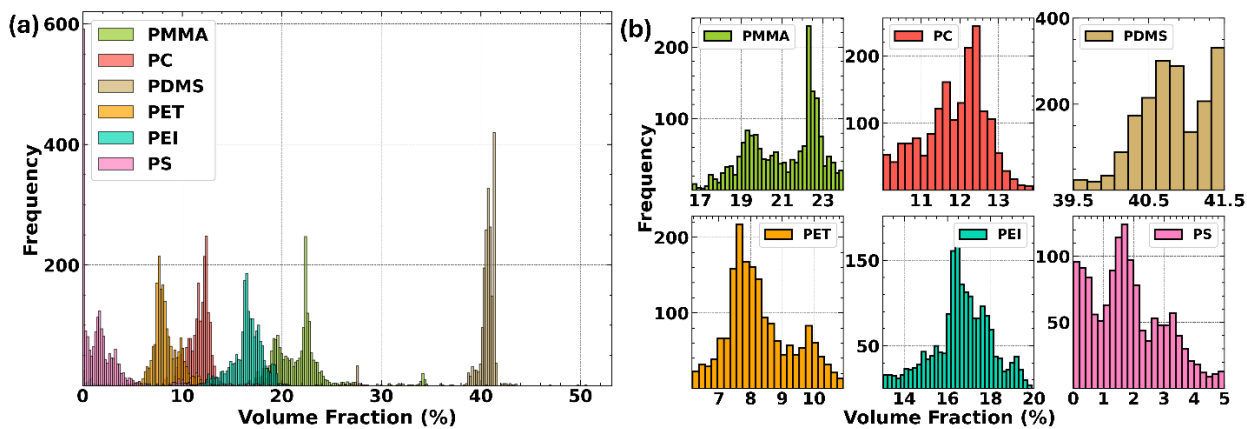

*Figure 8: Algorithm Stability and Precision under the Logarithmic Mixing Topology.* Results from the sensitivity study for the multi-component mixture (19%: 13%: 41%: 8%: 18%: 1% PMMA-PC-PDMS-PET-PEI-PS) using gradient descent across 2000 randomized initial conditions. The logarithmic mixing rule (representing statistical/random grain topology) is applied for the volume fraction reconstruction. Panel *(a)* shows combined histograms of all components across the full volume fraction spectrum and panel *(b)* illustrates the individual component distributions shown in separate frames.

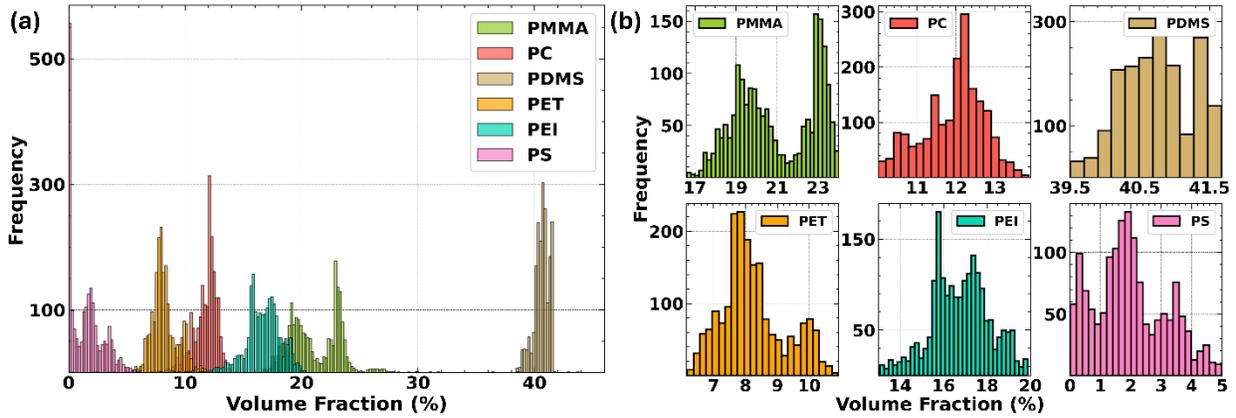

*Figure 9: Algorithm Stability and Precision under the Cubic Mixing Topology.* Results from the sensitivity study for the multi-component mixture (19%: 13%: 41%: 8%: 18%: 1% PMMA-PC-PDMS-PET-PEI-PS) using gradient descent across 2000 randomized initial conditions. The cubic mixing rule (representing co-continuous network topology) is applied for the volume fraction reconstruction. Panel *(a)* shows combined histograms of all components across the full volume fraction spectrum and panel *(b)* illustrates the individual component distributions shown in separate frames.

# Discussion

The results demonstrate that the proposed inverse reconstruction framework reliably establishes an inverse metrology standard by recovering both the broadband complex effective permittivity and constituent-level information of nonlinear mixtures using only a single IR extinction spectrum. The method's ability to consistently recover the effective material constant across various compositions (stage B) and its subsequent microstructure diagnosis (stage C) validates its generality for characterizing advanced optical composites.

In simulated mixtures, the algorithm accurately inferred both the number and identity of components, as well as their volume fractions, even under significant spectral overlap and nonlinear mixing. Crucially, the framework's ability to correctly reconstruct the known forward mixing model (e.g. matching the cubic inverse model to cubic input data) serves as a definitive validation of its diagnostic power. This performance, which maintained precision across inverted, logarithmic, and cubic mixing rules, highlights the framework's capability to distinguish between physically distinct interaction topologies (stratified/series, statistical/random grain, and co-continuous networks, respectively). Furthermore, reconstructions remained highly stable under randomized initializations, indicating a well-behaved optimization landscape suitable for automated and unsupervised metrology.

A notable strength of the framework is that it requires no prior knowledge of the number of components or the mixing model to initiate reconstruction. Instead, the algorithm automatically infers both the constituent identities and the interaction model. The framework achieves this by simultaneously comparing the residuals of multiple, competing EMA topologies while enforcing physical constraints (such as volume fraction summing to unity and the causality of the complex permittivity). This formulation enables a non-destructive, optical diagnosis of the material's internal microstructure – a critical parameter for rational material design that is not accessible via conventional spectroscopy. Unlike conventional approaches, such as peak-fitting [26], Beer-Lambert analysis [12], or unsupervised machine learning [61], our method is grounded in dielectric theory and provides physically interpretable results (complex effective permittivity and EMA topology), making it well suited for both diagnostics and materials applications where composition is unknown or highly variable.

Nevertheless, the framework has limitations. The current implementation relies on the assumption of spherical geometry and isotropic scattering (as rigorously defined by Mie theory), which limits its direct applicability to samples requiring different geometric forward models. Moreover, as demonstrated in the sensitivity analysis, the accuracy of volume fraction estimates is reduced for low-abundance components whose minimal spectral signals are overshadowed by strongly absorbing species. These challenges point to natural extensions, including support for alternative forward models, such as those for layered media (e.g. Fresnel-based models) or anisotropic scatterers. It may also be useful to integrate sparse or compressive sensing frameworks to enhance the quantification of low-abundance components and further improve computational efficiency.

The algorithm's performance scales sublinearly with mixture complexity. Increasing the number of components minimally expands the required basis set due to the sparsity of spectral features in the IR. Computational complexity is dominated by the robust gradient-based optimization in the mixture deconvolution stage (stage C), which performs well even in six-component cases, avoiding local minima and exhibiting low sensitivity to initialization. In contrast, grid search methods struggled with scalability due to the higher dimensionality of the parameter space, though they were retained in validation for completeness. Reconstruction times (typically 5-20 minutes on a standard desktop machine e.g., Intel i5-6500U CPU with 16 GB RAM) offer a practical balance of accuracy and computational efficiency for research and prototyping.

In conclusion, this work establishes a physics-guided inverse reconstruction method for inverse metrology in complex composites. By successfully resolving both the intrinsic effective permittivity and the underlying microstructure topology, the framework provides the necessary, physically grounded data to transition from empirical observation to the rational design and quantitative characterization of nonlinear optical materials.

# Conclusion

This work establishes a new inverse metrology standard for the characterization of complex, heterogeneous optical materials. We presented a physics-guided inverse reconstruction framework that successfully extracts the broadband complex effective permittivity and constituent composition of nonlinear, scattering mixtures directly from a single IR extinction spectrum.

By integrating rigorous Mie scattering theory, dispersion-consistent Lorentz oscillator models, and a generalized suite of nonlinear dielectric mixing rules (EMAs), the two-stage method achieves a comprehensive characterization: it recovers the dispersion characteristics of the effective medium, the identities and volume fractions of its underlying components, and, critically, performs a non-destructive diagnosis of the material's underlying microstructure topology.

The framework was rigorously validated on a large matrix of synthetic polymer mixtures across inverted, logarithmic, and cubic mixing regimes. The systematic agreement between the best-fit EMA and the known forward model serves as a definitive proof that the method can reliably distinguish between distinct physical topologies (stratified, random grain, and co-continuous). This topology diagnosis is a central and novel output that allows material scientists to link fabrication parameters directly to the observed macroscopic optical function.

Unlike conventional spectral analysis methods that rely on calibration, peak fitting, or assumed linearity, this approach is entirely physics-based, model-consistent, and highly interpretable, requiring no prior knowledge of the mixture components or their interaction mechanism. By providing both the scattering-immune effective permittivity for forward design applications and a physics-based diagnosis of the microstructure, the framework provides a computational foundation for quantitative hyperspectral imaging, soft-matter characterization, and the rational design of functional photonic composites, establishing a generalizable methodology for high-precision optical analysis of complex heterogeneous systems.

# Author Contributions

P.A. independently conceived the project, implemented the computational methods, conducted all simulations, and authored the manuscript.

# Funding

This research received no external funding.

# Conflict of Interest

The author declares no competing interest.

## Data Availability

All data used in this study are drawn from freely available public sources (see references where applicable). The computational framework relies on the open-source Python package PyMieScatt [56], and all analyses can be reproduced using this software. Detailed step-by-step instructions for reproducing the results are provided in the Supporting Information.